\documentclass[aip,apl,twocolumn,reprint,superscriptaddress]{revtex4-1}
\usepackage{graphicx}
\usepackage{epstopdf}

\usepackage{dcolumn}
\usepackage{bm}
\usepackage{amsmath}

\begin{document}

\title{Superconducting molybdenum-rhenium electrodes for single-molecule transport studies}
\author{R. Gaudenzi}
\altaffiliation{These authors contributed equally.}

\affiliation{Kavli Institute of Nanoscience, Delft University of Technology, 2600 GA, The Netherlands}
\author{J. O. Island}
\altaffiliation{These authors contributed equally.}
\affiliation{Kavli Institute of Nanoscience, Delft University of Technology, 2600 GA, The Netherlands}

\author{J. de Bruijckere}
\affiliation{Kavli Institute of Nanoscience, Delft University of Technology, 2600 GA, The Netherlands}
\author{E. Burzur\'{\i}}
\affiliation{Kavli Institute of Nanoscience, Delft University of Technology, 2600 GA, The Netherlands}
\author{T. M. Klapwijk}
\affiliation{Kavli Institute of Nanoscience, Delft University of Technology, 2600 GA, The Netherlands}
\affiliation{Physics Department, Moscow State Pedagogical University, Moscow 119991, Russia}
\author{H. S. J. van der Zant}

\affiliation{Kavli Institute of Nanoscience, Delft University of Technology, 2600 GA, The Netherlands}

\date{\today}

\begin{abstract}
\textbf{We demonstrate that electronic transport through single molecules or molecular ensembles, commonly based on gold (Au) electrodes, can be extended to superconducting electrodes by combining gold with molybdenum-rhenium (MoRe). This combination induces proximity-effect superconductivity in the gold to temperatures of at least 4.6 Kelvin and magnetic fields of 6 Tesla, improving on previously reported aluminum based superconducting nanojunctions. As a proof of concept, we show three-terminal superconductive transport measurements through an individual Fe$_4$ single-molecule magnet.}
\end{abstract}

\maketitle
\hfill 
\break
\indent Recent advances in nanostructure fabrication have made possible to couple superconductivity (SC) with confined systems of electrons. From this interaction, interesting phenomena like Andreev reflections \cite{Blonder1982} and Yu-Shiba-Rusinov states \cite{Soda1967, Shiba1969, Shiba1968, Rusinov1969} emerge where SC can be used alternatively as a probe to characterize the mesoscopic system \cite{Scheer1997} or as a tool to influence it \cite{Paaske2010, Paaske2011, Pascual2011}. When the confined system is an individual molecule or a nanocrystal, additional internal degrees of freedom like spin and vibrations are predicted to have an effect on the Cooper pair transport. For instance, supercurrent can be employed as a probe for isotropic and anisotropic spinful molecules \cite{Sadovskyy2011, Lee2008}. \\
\indent So far, only a handful of studies have investigated superconducting transport through individual molecules. Two recent examples are scanning tunneling microscopy studies using lead \cite{Pascual2011, Pascual2013} and two-terminal devices using tungsten \cite{Kasumov2005}. However, due to the absence of a gate, these studies are limited to the off-resonant transport regime and a single fixed charge state. Further studies, involving a combination of resonant transport and SC, are based on electromigrated gold break junctions \cite{Wernsdorfer2009}. These junctions are equipped with a gate electrode in close proximity to the molecule thereby forming a single-molecule transistor. Due to the difficulty of electromigrating materials other than gold, SC is typically induced by proximity to a superconducting material like aluminum \cite{Scheer2001}. The small superconducting gap of aluminum ($\Delta \approx 0.18$ meV, $T_\text{c} \approx 1.2 $ K, $B_\text{c} \approx 10 $ mT), however, limits the range of operation in magnetic field and temperature. In particular, the conditions for attaining the intermediate coupling transport regime ($\Gamma \sim \Delta \sim k_\text{B}T_{\text K}$) restrict the range of molecular couplings $\Gamma$ and Kondo energy scales $k_\text{B}T_{\text K}$. As a consequence limited room is left for the investigation of this intriguing regime where single-electron and many-body physics are directly competing \cite{Buitelaar2002, Defranceschi2010, Paaske2010, Paaske2011}. \\
\indent In this letter, we present a three-terminal hybrid electromigrated break junction, a SN-I-NS junction, that employs molybdenum-rhenium \cite{Talvacchio1986}  (MoRe) as superconducting material (S) and gold as normal metal (N). Gold allows for the creation of nanogaps (I) by electromigration and is commonly used for contacting single molecules due to its nearly ideal Fermi gas-like density of states (DOS), as well as inertness and compatibility with organic ligand terminations. When in contact with MoRe (60/40 alloy, $\Delta_\text{BCS} \approx 1.4$ meV, $\xi \approx 20$ nm \cite{Talvacchio1986}), we find that the gold junction exhibits a proximitized gap of about 0.7 meV. We characterize transport through these hybrid electromigrated junctions as a function of temperature and magnetic field. We demonstrate superconducting behavior up to at least a temperature of 4.6 K and a magnetic field of 6T. We show preliminary transport measurements resulting from coupling a fabricated SC junction to an individual Fe$_4$ single molecule magnet (SMM). \\ 
\indent The fabrication of the three-terminal device follows the procedure by Osorio \emph{et al.} \cite{Osorio2007} and only the relevant differences are described here. Conventional e-beam lithography and evaporation techniques are employed to fabricate the nanostructure. A scanning electron microscope (SEM) image of the device is shown in Fig. 1(a) and a corresponding side-view schematics is shown in Fig. 1(b). The stack consists of a 75 nm gold-palladium (AuPd) gate coated with 7 nm of atomic layer deposition-grown aluminum oxide (Al$_2$O$_3$) on top of which the thin gold wire is deposited. Two  MoRe superconducting contacts (110 nm-thick) partially overlap the gold wire, leaving a narrow, rectangular portion uncovered. This 260 nm-long and 100 nm-wide bridge forms the nanowire in which a nanogap is subsequently produced by room-temperature electromigration \cite{Park1999} and self-breaking \cite{Osorio2007}. In this method, a real-time feedback-controlled current of electrons is passed through the nanowire and used to displace the gold atoms (for the electromigration curve of a characteristic device see the Supplementary Material\cite{SuppInfo}). This process allows the formation of the SN-I-NS junctions, where the vacuum nanogap corresponds to the insulator sandwiched between the two gold portions of the normal wire and the MoRe superconducting patches. In the inset of Fig. 1(a) a SEM image of an electromigrated nanowire is shown. \\
\begin{figure}[t]
\includegraphics[width=0.99\columnwidth]{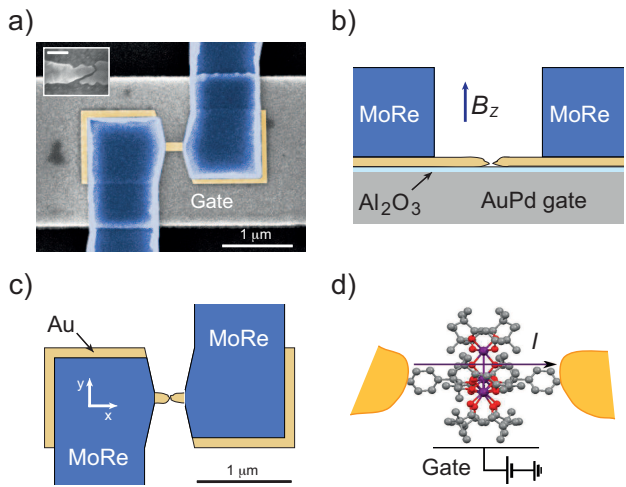}
  \caption{
  \textbf{The three-terminal hybrid MoRe-Au superconducting nanojunction.}
  (a) Scanning electron microscope micrograph of a three-terminal superconducting SNS junction (false colors) before electromigration. The two MoRe patches (purple), acting as source  and drain superconducting reservoirs, are in contact with the Au nanoribbon (yellow). The narrow part of the nanoribbon forms the nanowire to be electromigrated. A micrograph of an electromigrated junction is shown in the inset (100 nm scale bar). The z-axis is along the out-of-plane direction. (b) Side view schematics of an electromigrated junction. (c) Top view of (b). The x and y-axis are indicated. (d) Ideal arrangement of Fe$_4$ molecule between source and drain electrodes forming the three-terminal superconducting molecular transistor.}
\label{Fig_1}
\end{figure}
\indent The electromigrated SN-I-NS junctions are cooled down in a dilution fridge ($T \approx 20$ mK) equipped with a vector magnet. Temperature and magnetic field measurements are performed in a two-probe voltage-bias scheme, i.e. by applying a source-drain DC bias voltage ($V$) and recording the resulting current ($I$). The differential conductance spectrum $\text{d}I/\text{d}V$ versus $V$ is then obtained by taking the numerical derivative. A three-terminal measurement as a function of gate voltage ($V_\text{gate}$) and bias voltage is carried out to check for the absence of any gating and/or Coulomb blockade effect, see Supplementary Material\cite{SuppInfo}.  \\
\indent In Fig. 2(a) $\text{d}I/\text{d}V$ spectra as a function of temperature are shown. The low-temperature $\text{d}I/\text{d}V$ trace ($T = 100$ mK $ \ll T_c$) shows a V-shaped dip between two symmetric peaks at bias voltages $2V_\text{gap} = \pm 1.4 $mV. At higher biases, the conductance smoothly decreases to a plateau value, which we interpret as the normal state resistance regime. Increasing the temperature up to about 1.2 K is seen to only slightly affect the conductance at low voltage. In contrast, an increase in temperature from 2 K up to 3.1 K and further, results in a softening of the dip and a lowering of the two peaks, leaving the higher bias conductance unchanged throughout. Prominently, a residual dip is maintained up to the highest measured temperature of 4.6 K. \\
\indent The presence of a reduced gap-like structure in bias voltage with characteristic energy $E_\text{gap} < \Delta_\text{BCS}$ is a well-known signature of proximity-induced superconductivity \cite{Klapwijk2004}. This effect has been already experimentally observed, among others, in \citet{Gueron1996, Scheer2001} and theoretically investigated in detail by several authors \cite{Golubov1995, Belzig1996}. In these previous experiments as well as in the one discussed here, the superconducting coherence length $\xi$ compares with the characteristic lengths of the gold normal metal portion as $\xi, L \gg l_e \gg \lambda_{\text{F}}$, where $L$ is the length of the bridge, $l_e$ and $\lambda_{\text{F}}$ are the elastic scattering length and Fermi wavelength respectively. This situation is called the quasi-classical diffusive limit for which the theory has been formulated by Usadel \cite{Usadel1970}. A spatially-dependent density of states along with a reduced gap $E_\text{gap} (L) < \Delta$ arises from an application of the model to a N film of finite length L connected to a superconductor. Within this framework, the peaks and the dip in the $\text{d}I/\text{d}V$ spectra observed in this experiment result from the convolution of the superconducting peaks and the reduced gap in the proximity-induced DOS at the two N-I interfaces. \\
\indent We also investigate the persistence of SC upon application of an external magnetic field for different spatial directions. Fig. 2(b) shows the differential conductance spectra as a function of a field along the z-axis, i.e. perpendicular to the plane of the nanostructure (see Fig. 1(c)). A gradual decrease of the characteristic features is observed up to 1T. For higher magnetic field values a further decrease is accompanied by a complete suppression of the peaks at $\pm 1.4$ mV. A dip is present at the highest $B$-field value of 6 T signalling the presence of a residual superconducting DOS. Measurements with equivalent magnetic field intensities but along the y-axis, i.e. in-plane and perpendicular to the transport direction, are performed and the results displayed in Fig. 2(c). The softening of the dip and the coherence peaks for increasing magnetic fields is also observed. However, the spectra for the y-axis field maintain stronger superconducting features as compared to those of Fig. 2(b) for corresponding magnetic field values. Equivalently, the magnetic field $B_y$ acts comparatively weaker than $B_z$ in suppressing the proximity-effect SC. In analogy with the temperature-dependent measurements, we note that the high-bias regions ($eV > 2E_\text{gap}$) of the spectra are not affected by variations of the magnetic field.\\
\begin{figure*}[t]
\includegraphics[width=2.2\columnwidth]{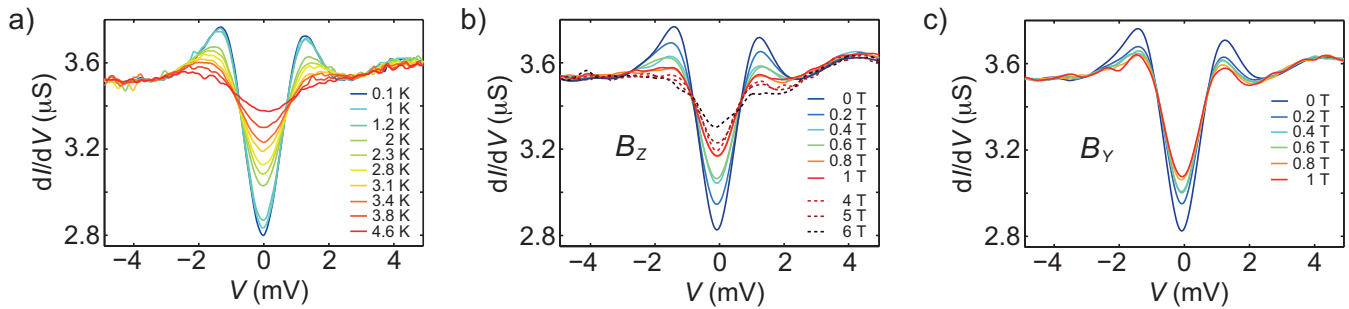}
  \caption{
  \textbf{Temperature and magnetic field voltage-bias characterization of the superconductivity.}
  (color online)
  (a) Differential conductance spectra measured as a function of  temperature ranging from 100 mK to 4.6 K. The characteristic gapped structure of the superconductive DOS persists up to above liquid-He temperature.  
  (b) Differential conductance traces measured as a function of the magnetic field along the z-axis at base temperature $T\approx 22$ mK. The dashed lines indicate the high magnetic field measurements. The signature of the superconducting gap is evident up to above a magnetic field $B_z = 6$ T.  (c) Same as (b) but with the magnetic field pointing along the y-axis. The solid lines indicate magnetic fields ranging from $B_y = 0$ T to $B_y = 1$ T. The characteristic peaks and the gap softens comparatively slower than in (b). Note that the vector magnet that we employed is limited to a magnetic field of 1 T along the y-axis.}
\label{Fig_2}
\end{figure*}
\indent The experimental magnetic field dependences can also be qualitatively explained within the diffusive Usadel framework. As shown in Belzig \emph{et al.} \cite{Belzig1996}, the applied magnetic field can be incorporated into an effective pair breaking rate $\Gamma_\text{eff}$ that affects the magnitude of the coherence peaks and the reduced gap energy. This pair breaking mechanism is proportional to the intensity of the magnetic field vector, $|B|$, as well as the dimension of the nanowire transverse to it, $W$, ($ \Gamma_\text{eff} \sim B^2 W^2$). In the present situation, the transverse directions corresponding to the magnetic fields $B_z$ and $B_y$ are the nanowire width and thickness, respectively. This would result in a stronger pair breaking effect along the z-axis as compared to the y-axis ($\Gamma^z_\text{eff}/\Gamma^y_\text{eff}\sim 100$), qualitatively consistent with the experimental observations (for an additional sample see the Supplementary Material\cite{SuppInfo}). We note that the persistence to high-magnetic fields can be partially ascribed to junction shape and/or geometry effects. \\ 
\indent Envisioning the use of our hybrid junctions as a superconducting molecular transistor, we present here preliminary results obtained from coupling an individual Fe$_4$ single-molecule magnet \cite{Accorsi2006} (SMM) to superconducting leads (schematically shown in Fig. 1(d)).  Fig. 3(a) displays the differential conductance map of an individual Fe$_4$-SMM as a function of gate and bias voltages for an external magnetic field $B = 0$ T. The standard features of sequential electron tunneling and Coulomb-blockade are seen. Each of the two low-conductance regions on either side of the charge degeneracy point ($V_\text{gate} \approx 2.5$ V, $V \approx 0$ V) corresponds to a stable charge state. Within these regions, the dip and the horizontal lines of increased conductance centered around zero-bias (black arrows) signal the expected SC density of states of the two leads. At the degeneracy point, the superconducting gap-like structure is lifted and a significant increase in zero-bias conductance occurs. In order to compare these observations with those on bare junctions, the differential conductance was measured as a function of magnetic field $B_z$, Fig. 3(b), at $V_\text{gate} = 1.95$ V, far into the off-resonant regime (dashed line in Fig. 3(a)). A reduced gap $2E_\text{gap} \approx 0.7$ meV appears. Gradual suppression of the superconducting features takes place from zero magnetic field to about 0.6 T, leaving a residual gap structure weakly evolving from 0.6 T to 1T. In the inset of Fig. 3(b) the differential conductance map from which the spectra are extracted is shown. The magnetic field ranges from -1 T to +1T. The smoothing of the superconducting features is symmetric for negative and positive field values.  \\
\begin{figure}[t]
\includegraphics[width=0.99\columnwidth]{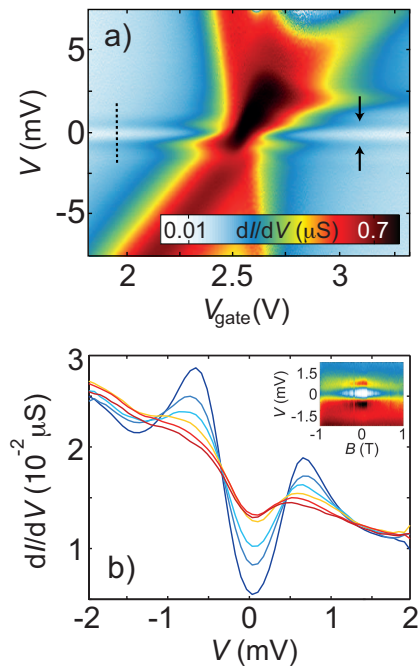}
  \caption{
  \textbf{The superconducting single-molecule transistor.}
  (color online) (a) Differential conductance map as a function of gate and bias voltages measured at $B = 0$ T and $T = 0.6$ K. Superconductivity and Coulomb blockaded transport superimpose in the two stable charged states. The horizontal lines of increased conductance (marked by black arrows) and the low-bias dip indicate the superconducting DOS of the leads. At the charge degeneracy point ($V_\text{g} \approx 2.5$ V) the superconducting gap is lifted and the conductance greatly increases. (b) Differential conductance spectra as a function of magnetic field and bias voltage at fixed gate voltage $V_\text{gate} = 1.95$ V (dashed line in the left stable charge state of (a)). The spectra are extracted from the map in the inset, starting from $B = 0$ T (blue line) to $B = 1$ T (red line) at a regular spacing $\Delta B = 0.2$ T. A weak trace of the gapped DOS is still visible at $B = 1$T.}
\label{Fig_3}
\end{figure}
\indent In the present example, the charging energy $U \geq100$ meV and the tunneling rate $\Gamma \approx 1$ meV \footnote{The value of $\Gamma$ is extracted from the FWHM of the lorentzian fit of the Coulomb peak at a magnetic field of 8T, in order to minimize the influence of superconductivity on transport. The lower bound value for $U$ is estimated from the full $V$ vs $V_{gate}$ conductance map (See also the Supplementary Material\cite{SuppInfo}).}, characteristic energies of single-electron transport, are related to $E_\text{gap}$ by $U \gg \Gamma \gtrsim E_\text{gap} $. The first condition, $U \gg \Gamma$, guarantees Coulomb blockade and single-electron-transistor behavior \cite{Thijssen2008}. The second condition, $\Gamma \gtrsim E_\text{gap} $, allows for the off-resonant inelastic quasiparticle tunneling and would theoretically enable the on-resonant transport of both single electrons and Cooper pairs \cite{Glazman1989}. The off-resonant transport and the strong increase in zero-bias conductance observed in Fig. 3(a) are consistent with this picture and will be the subject of further study. \\ 
\indent We have presented a three-terminal hybrid electromigrated break junction with high-critical field superconducting electrodes for single-molecule studies. In this SN-I-NS junction, superconductivity is induced in the gold by proximitizing it with MoRe. Gold as a normal metal allows for the creation of nanogaps by controlled electromigration and preserves the advantage of molecule-gold chemistry. The use of MoRe as a superconductor guarantees an induced gap larger than the previously reported Al-based designs. We characterize induced superconductivity as a function of temperature and magnetic field intensity and direction and demonstrate superconducting behavior up to 4.6 K and a critical magnetic field of 6T. We finally show preliminary transport measurements through an individual Fe$_4$ single molecule magnet. Coexistence of Coulomb blockade and superconducting transport is observed. Low-bias on- and off-resonant conductance behavior suggests that, owing to the relatively high gap energy $E_\text{gap}$, the condition $U \gg \Gamma \gtrsim E_\text{gap} $ for intermediate coupling transport is satisfied. This intermediate coupling transport regime appears to be promising for investigating the interaction between confined electrons and superconductivity. Moreover, it constitutes the prerequisite - with the additional condition $k_\text{B}T_{\text K} \sim E_\text{gap}$ - for the investigation of the interplay between Kondo screening and superconducting pairing. \\
\indent We acknowledge A. Cornia for the chemical synthesis of the Fe$_4$ single molecule magnet and A. Holovchenko for assistance in obtaining the SEM micrographs. This work was supported by an advanced ERC grant (Mols@Mols). We also acknowledge financial support by the Dutch Organization for Fundamental research (NWO/FOM).  T. M. Klapwijk acknowledges financial support from the Ministry of Science and Education of Russia under Contract No.14.B25.31.0007 and from the European Research Council Advanced Grant No. 339306 (METIQUM). 

\bibliographystyle{apsrev4-1}
\bibliography{Biblio}

\end{document}